\documentclass[a4paper]{jpconf}
\usepackage{graphicx}
\usepackage{bm}
\usepackage{citesort}
\begin{document}
\title{Testing collinear factorization and nuclear parton distributions with pA collisions at the LHC}

\author{\underline{Paloma Quiroga-Arias}$^{1}$, Jos\'e Guilherme Milhano$^{2,3}$ and Urs Achim Wiedemann$^{3}$}

\address{$^1$ Departamento de F\'isica de Part\'iculas and IGFAE, Universidade de Santiago de Compostela 15706 Santiago de Compostela, Spain\\
$^2$ CENTRA, Departamento de F\'isica, Instituto Superior T\'ecnico (IST),
Av. Rovisco Pais 1, P-1049-001 Lisboa, Portugal\\
$^3$ Physics Department, 
    Theory Unit, CERN,
    CH-1211 Gen\`eve 23, Switzerland}


\ead{pquiroga@fpaxp1.usc.es}

\begin{abstract}
Global perturbative QCD analyses, based on large data sets from electron-proton and hadron collider experiments, provide tight constraints on the parton distribution function (PDF) in the proton. The extension of these analyses to nuclear parton distributions (nPDF) has attracted much interest in recent years. nPDFs are needed as benchmarks for the characterization of hot QCD matter in nucleus-nucleus collisions, and attract further interest since they may show novel signatures of non- linear density-dependent QCD evolution. However, it is not known from first principles whether the factorization of long-range phenomena into process-independent parton distribution, which underlies global PDF extractions for the proton, extends to nuclear effects. As a consequence, assessing the reliability of nPDFs for benchmark calculations goes beyond testing the numerical accuracy of their extraction and requires phenomenological tests of the factorization assumption. Here we argue that a proton-nucleus collision program at the LHC would provide a set of measurements allowing for unprecedented tests of the factorization assumption underlying global nPDF fits.
\end{abstract}

\section{Introduction}
Parton distribution functions (PDFs), which define the
flux of quarks and gluons ($i = q, g$) in hadrons $f_{i/h}(x,Q^2)$, play a central role in the study of
high energy collisions involving hadronic projectiles. For protons, sets of collinearly factorized
universal PDFs are obtained in global pQCD analyses based on data from DIS and DY production, 
as well as W/Z and jet production at hadron colliders. These data provide tight constraints
on PDFs over logarithmically wide ranges in $Q^2$ and $x$, and have allowed precision 
testing of linear pQCD evolution.
However, our understanding of parton distribution functions in nuclei of nucleon number $A$, $f_{i/A}(x,Q^2)$, is much less mature. Knowledge of nuclear parton distribution functions
(nPDFs) is important  in heavy ion collisions at RHIC and at the LHC for a quantitative control
of hard processes, which are employed as probes of dense QCD matter.
Characterizing nPDFs is also of great interest in its own right, 
since the nuclear environment is expected to enhance parton density-dependent effects, which
can reveal qualitatively novel, non-linear features of QCD evolution.
 
Paralleling the determination of proton PDFs, several global
QCD analyses of nPDFs have been made within the last decade~\cite{Eskola:2009uj,Eskola:2008ca,Eskola:1998df,deFlorian:2003qf,Hirai:2007sx} based, up until recently, solely on fixed-target nuclear DIS and DY data which, compared to the data constraining proton PDFs, are of lower precision and cover a much more limited range of $Q^2$ and $x$. Such limitation manifests in a particularly poor gluon distribution function, since it can only be obtained from logarithmic $Q^2$-evolution, for which a wide $Q^2$-range is mandatory (see~\cite{LAMONT} for future prospects in DIS on nuclei).


Moreover, in contrast to the theoretical basis for global analyses of proton PDFs, the separability of nuclear effects into process-independent nPDFs and process-dependent but A-independent hard processes is not established within the framework of collinear factorized QCD. 


In view of the importance of nPDFs for characterizing 
benchmark processes in heavy ion collisions, it is thus desirable to look for both stringent
phenomenological tests of the working assumption of global nPDF fits and experimental observables that can be used to constrain nPDFs. To this end, we argue that a program of hadron-nucleus collisions at the LHC would provide for such tests with unprecedented quality~\cite{QuirogaArias:2010wh}. 


\section{Collinearly factorized single inclusive approach to particle production}

To improve on the insufficiency of nuclear DIS and DY data, recent global nPDF analysis~\cite{Eskola:2008ca,Eskola:2009uj} have included for the first time 
data from inclusive high-$p_T$ hadron production in hadron-nucleus scattering measured at RHIC~\cite{Adler:2006wg,Adams:2006nd,Arsene:2004ux}. This is the observable we focus on in order to test collinear factorization and to determine nPDFs.

In the factorized QCD ansatz to hadron-nucleus collisions by construction the entire nuclear dependence of the cross section for hadro-production resides in the nPDF (see eq.(1) in~\cite{QuirogaArias:2010wh}).
%
%
It is customary to characterize nuclear effects in the parton distribution functions by the ratios 
\begin{equation}
  R_i^A(x,Q^2) \equiv	f_{i/A}(x,Q^2) \big/ f_{i/p}(x,Q^2)\, ,
  \label{Rpdf}
\end{equation}
which show characteristic deviations from unity in global nPDF analysis for all scales of $Q^2$ tested so far. These effects are typically referred to as nuclear shadowing ($x < 0.01$), anti-shadowing ($0.01 < x < 0.2$), EMC effect ($0.2 < x < 0.7$) and Fermi motion ($x > 0.7$).  

Nuclear effects on single inclusive hadron production are typically characterized 
by  the nuclear modification factor $R_{p\, A}^{h}$
\begin{equation}
 	R_{p\, A}^{h}(p_T, y) = \frac{d\sigma^{pA \to h+X}}{dp_T^2\, dy}  \Bigg /  
	N_{\rm coll}^{pA} \frac{d\sigma^{pp \to h+X}}{dp_T^2\, dy}\, .
	\label{RpA}
\end{equation}
%

 We calculate the single inclusive $\pi^0$ production at both RHIC and the LHC within the collinear factorization approach. All our calculations use LO PDFs from CTEQ6L~\cite{Pumplin:2005rh} with nuclear modifications EPS09LO~\cite{Eskola:2009uj} and the KKP fragmentation functions~\cite{Kniehl:2000fe}. When the collinear factorization ansatz is adopted, the $p_T$-dependence of $R_{p\, A}^{h}$ traces the $x$-dependence of nPDFs. Although the precise kinematics is complicated by the convolution of the
 distributions, the qualitative dependence of the Bjorken-x of partons inside the nucleus with $p_T$ of the final hadron and rapidity reads $x\sim p_Te^{-y}/\sqrt{s}$.

\subsection{From RHIC to the LHC: testing collinear factorization}

\begin{figure}[h]
\vspace{-1pc}
\begin{minipage}{17pc}
\includegraphics[width=17pc,height=10pc]{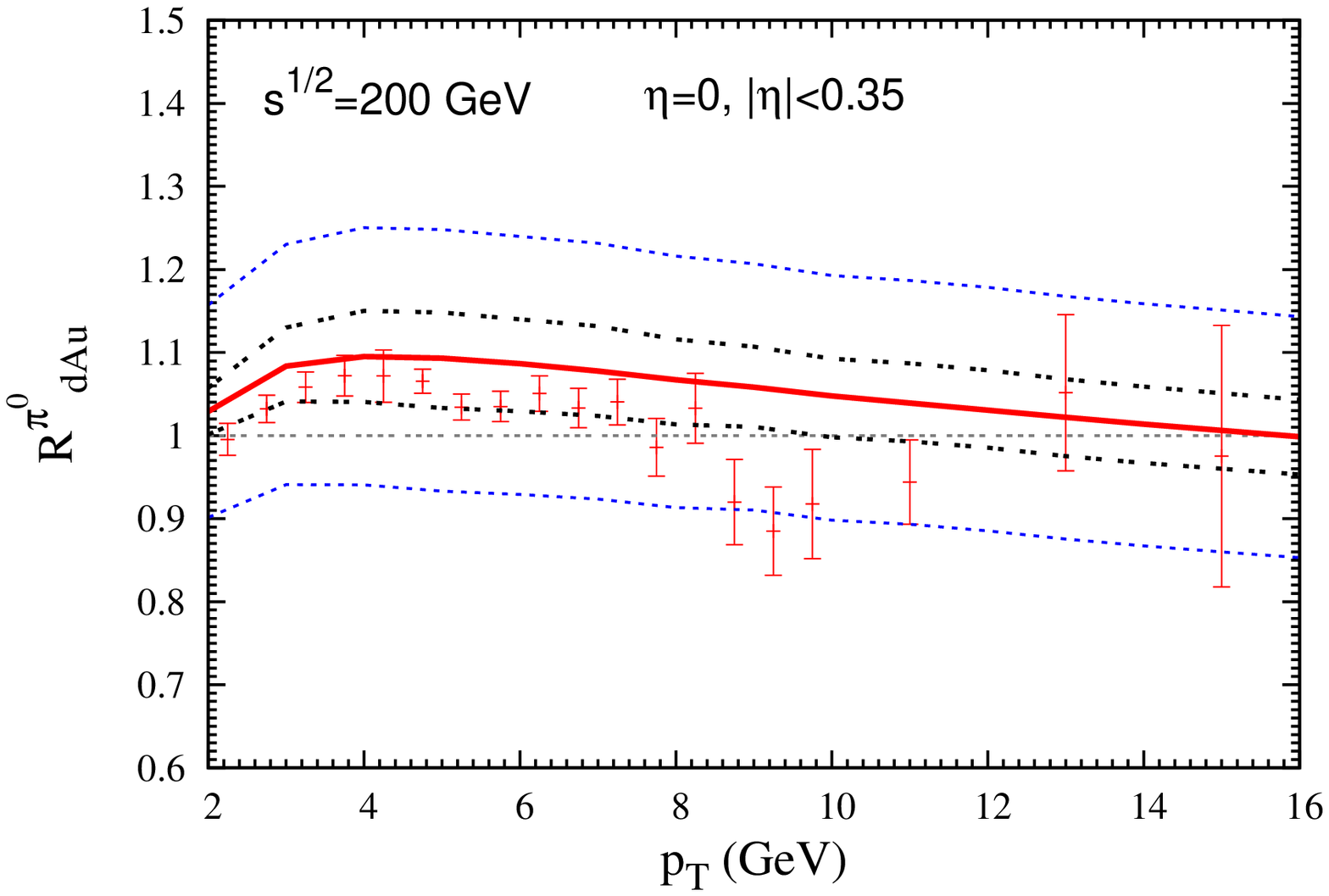}
\vspace{-2.8pc}\caption{\label{Fig1} Nuclear modification factor (\ref{RpA}) at RHIC mid-rapidity. Data points are PHENIX~\cite{Adler:2006wg}. Red line represents the EPS09 LO calculation. The dashed lines are the EPS09 uncertainty~\cite{Eskola:2009uj} (black) plus data normalization uncertainty (blue).}
\end{minipage}\hspace{2pc}%
\begin{minipage}{17pc}
\includegraphics[width=17pc,height=9.5pc]{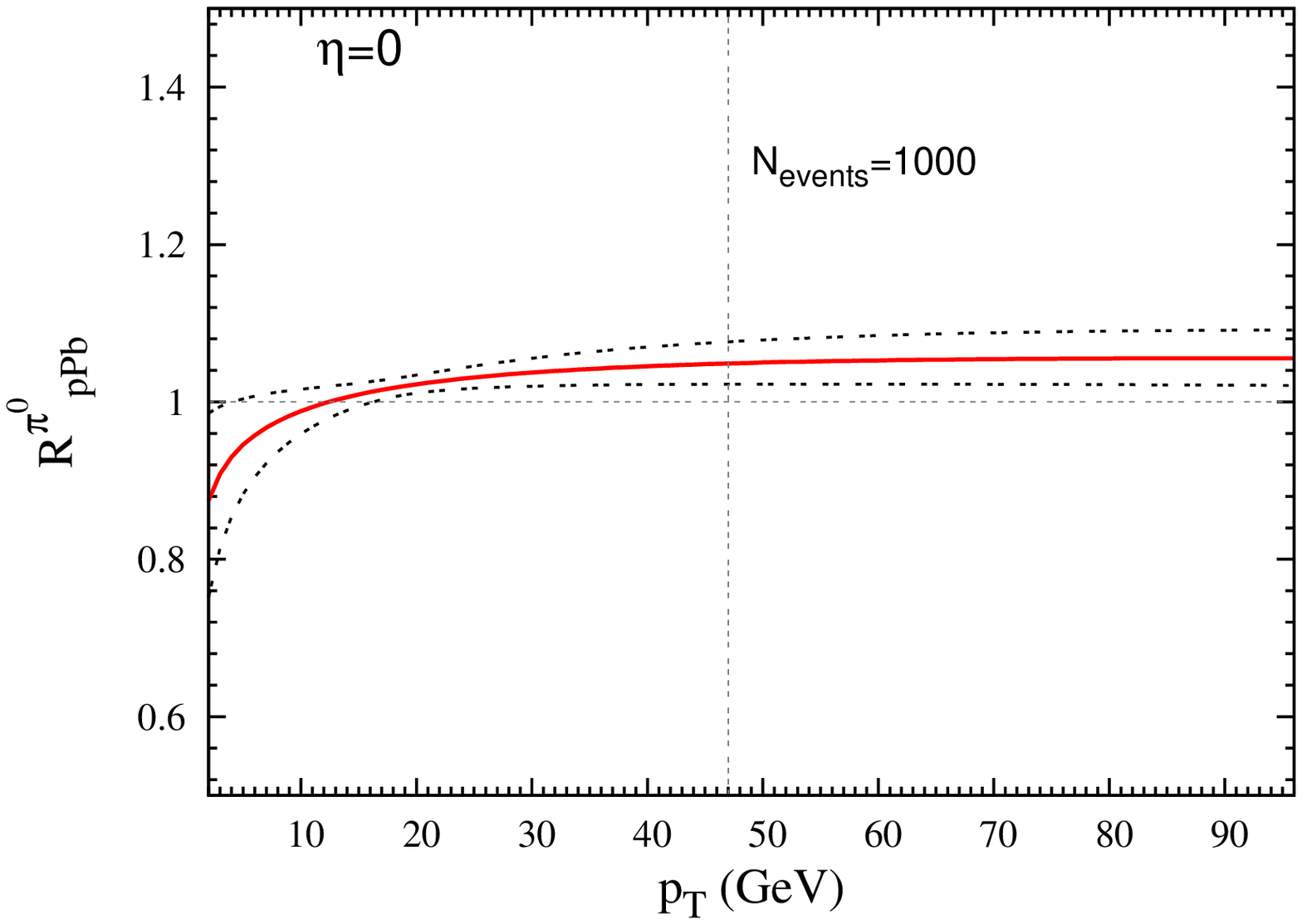}
\vspace{-2.2pc}\caption{\label{Fig2} EPS09 LO prediction for the nuclear modification factor of neutral pion single inclusive production in proton-lead collisions at LHC mid-rapidity. All LHC rapidities are in the center-of-mass (shift to lab $\Delta y\sim 0.47$).}
\end{minipage} 
\vspace{-1pc}
\end{figure}

Fig.~\ref{Fig1} shows the
nuclear modification factor $R_{d\, Au}^{\pi^0}$ for the production of neutral pions in $\sqrt{s_{\rm NN}} = 200$ GeV deuteron-gold collisions at RHIC, calculated within the factorized ansatz at leading order (LO). Results shown in Fig.~\ref{Fig1} are also 
consistent with the NLO-calculation of $R_{d\, Au}^{\pi^0}$ in \cite{Eskola:2009uj}.  The RHIC data~\cite{Adler:2006wg} in Fig.~\ref{Fig1} have 
 been used in constraining the nPDF analysis EPS09 \cite{Eskola:2009uj}
 but they were not employed in a closely related nPDF fit~\cite{Eskola:1998df}, which
 provides an equally satisfactory description of these RHIC data.  
Therefore, the agreement of data and calculation  in 
 Fig.~\ref{Fig1} is in support of collinear factorization: the enhancement of $R_{d\, Au}^{\pi^0}$ in the region around $p_T \simeq 4$ GeV  at mid-rapidity tests momentum fractions in the anti-shadowing region.

  
However, qualitatively different explanations of the $R_{d\, Au}^{\pi^0}$
 measured at RHIC are conceivable~\cite{AlbaceteHQ}. While the above calculation accounted for  $R_{d\, Au}^{\pi^0}(\eta=0)$ in terms of a 
nuclear modification of the {\it longitudinal} parton momentum distribution only, such nuclear effects can be accommodated within a variety of assumptions, namely initial state parton multiple scattering ($k_T$ broadening)~\cite{Zhang:2001ce} and non-linear low-$x$ QCD evolution~\cite{Albacete:2010bs}. 
 
Yet, RHIC data being inconclusive about the dynamical explanation of the nuclear modification of the spectrum, we argue that repeating the same measurement at the LHC at a much higher energy and in the much wider available $p_T$ range, will help to disentangle which of the theories is correct, or in which kinematical range each one is valid.


As one can see in Fig.~\ref{Fig2}, the shape of the nuclear modification factor that would be measured at LHC mid-rapidity, if the entire nuclear effect in pPb collisions can be factorized into nPDFs, is qualitatively different from that observed at RHIC: at 
more than 40 times higher center of mass energy, final state hadrons at the same transverse 
momentum test ${\cal{O}}$(40) times smaller momentum fractions $x_i$. Even tough a shift of the maximum of $R_{pPb}^{\pi^0}(\eta=0)$ to such high values of $p_T$ is a natural consequence in collinear factorization reflecting nuclear modifications of PDFs, it cannot be accounted for in terms of $k_T$-broadening, since such models predict a mild dependence of the spectrum with the center-of-mass energy~\cite{Zhang:2001ce}.

\begin{figure}[h]
\vspace{-5pc}
\includegraphics[width=36pc]{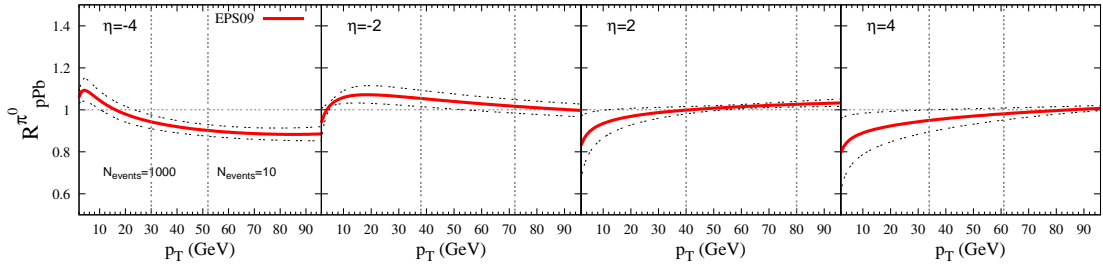}\vspace{-1.5pc}
\caption{\label{Fig3}Rapidity dependence of $R_{pPb}^{\pi^0}$ (\ref{RpA}) 
for $\sqrt{s_{\rm NN}}= 8.8$ TeV pPb (LHC). The different plots scan the dependence
from $y = -4$ (close to Pb projectile rapidity) up to $y = 4$ (close to proton projectile rapidity).
Labels indicate whether the nuclear modification originates mainly from the shadowing (S),
anti-shadowing (AS) or EMC regime. Vertical lines illustrate the rapidity-dependent $p_T$
range, which can be accessed experimentally with more than $N_{\rm events} = 1000$ (= 10)
per GeV-bin within one month of running at nominal luminosity.}
\vspace{-1.5pc}
\end{figure}

When the single-inclusive spectrum is calculated
for different values of rapidity the
results shown in figure Fig.~\ref{Fig3} are obtained, to learn that the rapidity dependence of $R^h_{pPb}$ allows one to scan the main qualitatively different ranges of standard nPDFs in an unprecedented way.

The behavior of the nuclear modification factor when saturation is taken
into account~\cite{Albacete:2010bs} is rather different: a strong suppression factor is found which is incompatible with any of the existing nuclear parton distribution sets. Thus, a measurement at the LHC which shows the rapidity dependent behavior in Fig.~\ref{Fig2} and~\ref{Fig3} would be in strong support of collinear factorization as opposed to initial state multiple scattering and CGC models.


\subsection{Constraining nPDFs with the LHC}
\begin{figure}[h]
\vspace{-1.5pc}
\begin{minipage}{20pc}
\includegraphics[width=20pc,height=11pc]{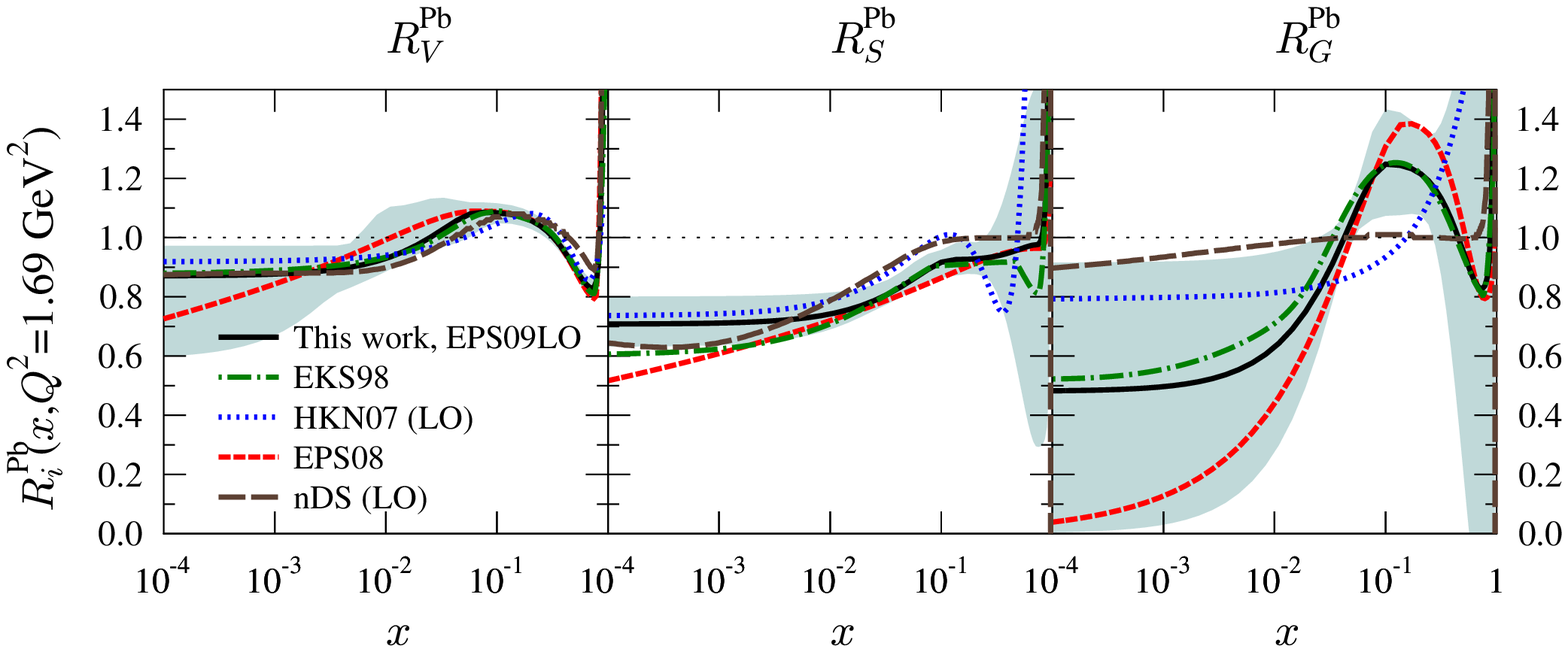}
\vspace{-1.5pc}
\caption{\label{Fig4}The nuclear modification factor~(\ref{Rpdf}) in lead $R^{Pb}_i$ for three different flavours as given by the available nPDF parametrization at LO.}
\end{minipage}\hspace{2pc}%
\begin{minipage}{16pc}
\includegraphics[width=16pc,height=11pc]{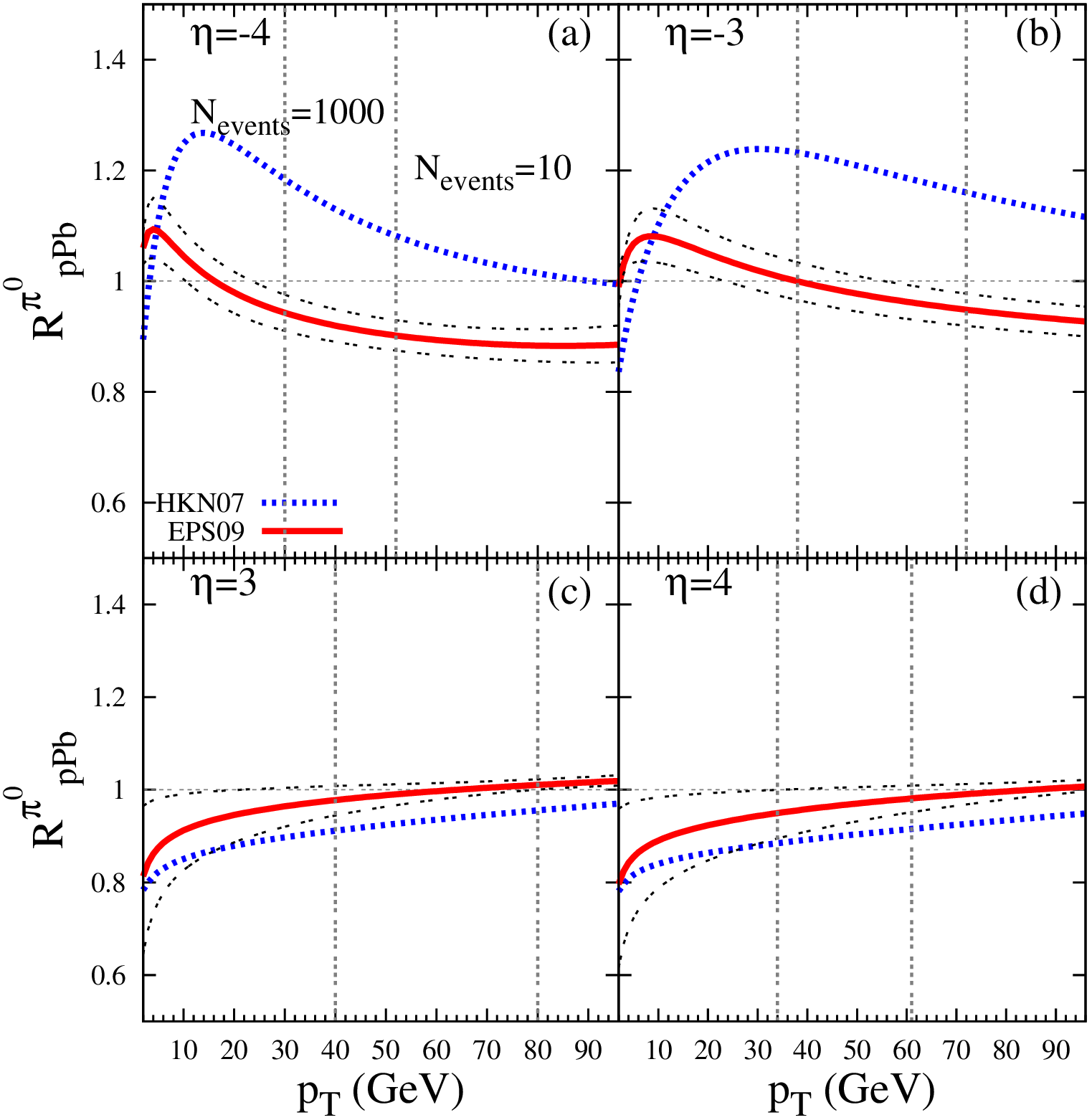}
\vspace{-1.5pc}
\caption{\label{Fig5}$R_{pPb}^{\pi^0}$ (\ref{RpA}) 
for $\sqrt{s_{\rm NN}}= 8.8$ TeV pPb (LHC), from two different
sets of nPDFs.}
\end{minipage} 
\vspace{-2pc}
\end{figure}

Despite the perspectives for qualitative tests of collinear factorization, current global analyses of nuclear parton distribution functions come with significant uncertainties as we can see in Fig.~\ref{Fig4}. We have compared in Fig.~\ref{Fig5} the nuclear modification factor
for two nPDF sets, which show marked differences: in contrast to EPS09 the gluon distribution of HKN07~\cite{Hirai:2007sx} does not show an 
anti-shadowing peak but turns for $x > 0.2$ from suppression to strong 
enhancement  at initial scale $Q^2 = 1\, {\rm GeV}^2$. Fig.~\ref{Fig5} thus illustrates that within the validity of a collinearly factorized approach, LHC data can resolve the qualitative differences between existing nPDF analyses and can
improve significantly and within a nominally perturbative regime on our knowledge of nuclear gluon distribution functions. 

\ack
We acknowledge support from MICINN (Spain) under project FPA2008-01177 and FPU grant; Xunta de Galicia (Conseller\'ia de Educaci\'on) and through grant PGIDIT07PXIB206126PR, the Spanish Consolider-
 Ingenio 2010 Programme CPAN (CSD2007-00042) and Marie Curie MEST-CT-2005-020238-EUROTHEPHY (PQA); and Funda\c c\~ao para a Ci\^encia e a Tecnologia (Portugal) under project CERN/FP/109356/2009 (JGM).

\section*{References}
\bibliography{bib}


\end{document}